\documentclass[usegraphicx]{mn2e}

\usepackage{amssymb}
\usepackage{mathptmx}

\newcommand{\Mpc}{\rm\thinspace Mpc}
\newcommand{\kpc}{\rm\thinspace kpc}

\newcommand{\km}{\rm\thinspace km}

\newcommand{\cm}{\rm\thinspace cm}
\newcommand{\s}{\rm\thinspace s}

\newcommand{\erg}{\rm\thinspace erg}

\newcommand{\ergpcmcu}{\hbox{$\erg\cm^{-3}\,$}}
\newcommand{\ergpcmsqps}{\hbox{$\erg\cm^{-2}\s^{-1}\,$}}

\newcommand{\ergps}{\hbox{$\erg\s^{-1}\,$}}

\newcommand{\kmps}{\hbox{$\km\s^{-1}\,$}}

\newcommand{\kmpspMpc}{\hbox{$\kmps\Mpc^{-1}$}}

\newcommand{\Lsun}{\hbox{$\rm\thinspace L_{\odot}$}}

\newcommand{\Zsun}{\hbox{$\thinspace \mathrm{Z}_{\odot}$}}

\begin{document}
\title[Fossil bubbles in the Perseus cluster] {Non-thermal X-rays, a
  high abundance ridge and fossil bubbles in the core of the Perseus
  cluster of galaxies}

\author[J.S. Sanders, A.C. Fabian and R.J.H.  Dunn] {J.S.
  Sanders\thanks{E-mail: jss@ast.cam.ac.uk},
  A.C. Fabian and R.J.H. Dunn\\
  Institute of Astronomy, Madingley Road, Cambridge CB3 0HA}

\maketitle

\begin{abstract}
  Using a deep \emph{Chandra} observation of the Perseus cluster of
  galaxies, we find a high-abundance shell 250~arcsec (93~kpc) from
  the central nucleus.  This ridge lies at the edge of the Perseus
  radio mini-halo. In addition we identify two H$\alpha$ filaments
  pointing towards this shell. We hypothesise that this ridge is the
  edge of a fossil radio bubble, formed by entrained enriched material
  lifted from the core of the cluster. There is a temperature jump
  outside the shell, but the pressure is continuous indicating a cold
  front.  A non-thermal component is mapped over the core of the
  cluster with a morphology similar to the mini-halo.  Its total
  luminosity is $4.8 \times 10^{43}\ergps$, extending in radius to
  $\sim 75 \kpc$. Assuming the non-thermal emission is the result of
  inverse Compton scattering of the CMB and infrared emission from
  NGC~1275, we map the magnetic field over the core of the cluster.
\end{abstract}

\begin{keywords}
  X-rays: galaxies --- galaxies: clusters: individual: Perseus ---
  intergalactic medium
\end{keywords}

\section{Introduction}
The Perseus cluster, Abell~426, has long been known to host
depressions in its X-ray surface brightness image. Fabian et al (1981)
and Branduardi-Raymont et al (1981) identified a hole in emission
around 80~arcsec north-west of the nucleus using the \emph{Einstein
  observatory}. Using \emph{ROSAT}, B\"ohringer et al (1993) found two
further inner depressions to the north-east and south-west of the
core, of size $\sim 0.5$~arcmin.  These holes in the X-ray emission
coincided in position with the radio lobes of the bright central radio
source 3C~84 (Pedlar et al 1990).  It is therefore probable that the
radio plasma has displaced the thermal X-ray emitting gas from the
bubbles.

Using the sub-arcsecond imaging capabilities of the \emph{Chandra
  observatory}, Fabian et al (2000) found that the X-ray bright rims
of the inner radio lobes are cooler than the surrounding gas. They
also identified a further outer X-ray hole to the south of the
nucleus. The two outer X-ray holes correspond in position to two
spurs in the low frequency radio emission (Fabian et al 2002). This
low frequency emission is probably due to a population of old
electrons, indicating that these depressions are two fossil radio lobes
which have detached themselves from the nucleus.

Deep \emph{Chandra} images of the cluster show there appears to be a
weak shock driven by the inner north-east radio bubble (Fabian et al
2003a).  Furthermore, evidence for ripples in the X-ray surface
brightness was found which may be waves driven in the intracluster
medium (ICM) by the expansion of the radio lobes. Detailed spectral
analysis of the data showed evidence that the gas was enriched around at
least two of the holes (Sanders et al 2004). This supports the
hypothesis that rising radio lobes lift and entrain high abundance
material from the core of the cluster (Churazov et al 2001).

In addition the cluster exhibits a large number of optical filaments
emitting in H$\alpha$ (Conselice, Gallagher and Wyse 2001). Many of
these filaments are radial, and remarkably straight. The morphology of
these filaments supports the argument that the gas in these clusters is
viscous and not turbulent (Fabian et al 2003b).

The Perseus cluster is at a redshift of 0.0183. We assume that $H_0 =
70 \kmpspMpc$; therefore 1 kpc corresponds to about 2.7 arcsec.

\section{Analysis}
\subsection{Temperature and abundance structure}
\label{sect:temp}
For this analysis we examined the 191-ks deep \emph{Chandra}
observation of the Perseus cluster. The data were processed using the
same prescription as given in Sanders et al (2004).

We selected regions in the cluster using the contour binning algorithm
of Sanders~(in preparation). The method takes an adaptively smoothed
X-ray image of the cluster, and uses it to define regions which have
similar surface brightness. Firstly the routine adaptively smooths an
X-ray image using a method called `accumulative smoothing'. This form
of adaptive smoothing is similar to that used by the \textsc{ftools}
routine \textsc{fadapt}.  It smooths using a top-hat circular kernel
which varies in size in order to have a minimum signal to noise within
the smoothing kernel ($S/N \sim n / \sqrt{n}$ if background is not
taken into account, where $n$ is the number of counts in the kernel).

The contour binning algorithm starts at the highest flux pixel in the
smoothed image. This pixel is added to the current bin. If the signal
to noise within this bin (again $\sim n / \sqrt{n}$, if background is
ignored, where $n$ is the total number of counts within the bin on the
unsmoothed image) is less than a threshold value, chosen here to be
100, then the neighbouring pixel closest in smoothed flux to the
starting pixel is added to the bin. This is repeated until the signal
to noise threshold is reached or there are no remaining neighbouring
pixels. Additionally, we apply a geometric constraint to pixels to be
added to a bin. We do not add pixels if they lie a distance of greater
than $f r$ away from the current centroid of the bin, where $r$ is the
radius a circle would have if it had the same area as the bin
currently does, and $f$ is a parameter which we choose to be 1.8. This
constraint ensures the bins do not become too elongated. Increasing
$f$ allows bins which are more elongated. When a bin is completed,
binning starts on the highest remaining pixel on the smoothed map.

When there are no remaining pixels, pixels in bins which contain a
signal to noise less than the threshold value are transferred to
neighbouring bins closest in smoothed flux. These low signal to noise
bins are created if they have no neighbouring pixels which could be
added.  A signal to noise of 100 yields bins which contain $>10^4$
counts.

Spectra were extracted from each region. Responses were generated
using a custom tool to add responses built using the \textsc{ciao}
\textsc{mkrmf} program, weighted relative to the number of counts in
each response region between 0.5 and 7~keV. Ancillary responses were
generated using the \textsc{mkwarf} tool. We extracted background
spectra from a blank sky observation, for each region, as was done in
Sanders et al (2004). Spectra were binned to contain a minimum of 20
counts per spectral channel. Each spectrum was then fitted with
\textsc{xspec} (Arnaud 1996) using a \textsc{mekal} model (Mewe,
Gronenschild \& van den Oord 1985; Liedahl, Osterheld \& Goldstein
1995) absorbed by a \textsc{phabs} model (Balucinska-Church \&
McCammon 1992). In the fits, the temperature, abundance (relative to
solar), normalisation and absorption were free. The redshift was set
to the mean redshift found in our previous analysis, 0.0169 (Sanders
et al 2004). We assumed the solar abundance ratios of Anders \&
Grevesse (1989) when fitting.

\begin{figure}
  \includegraphics[width=\columnwidth]{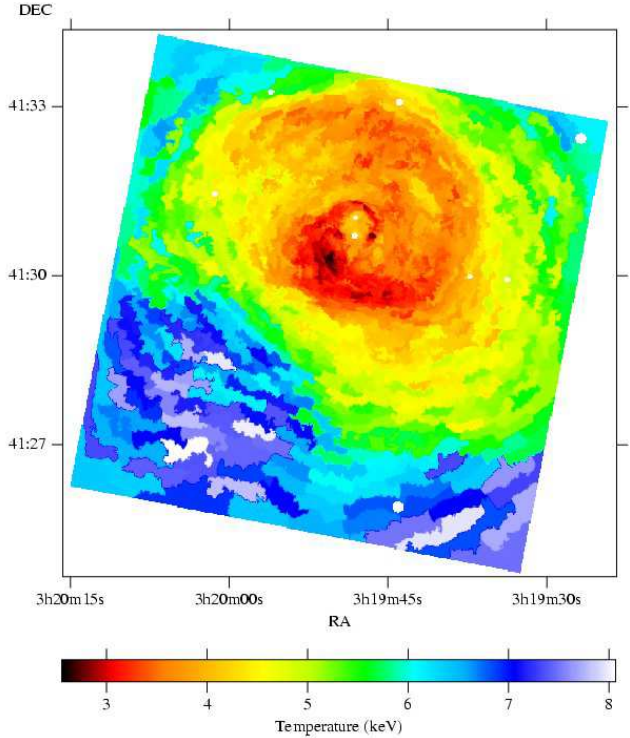}
  \caption{Temperature map of the cluster. The uncertainties on the
    temperatures range from 0.1~keV in the coolest region to 0.5~keV
    in the hottest regions. The white circles are excluded point
    sources, listed in Sanders et al (2004).}
\label{fig:temperature}
\end{figure}

\begin{figure}
  \includegraphics[width=\columnwidth]{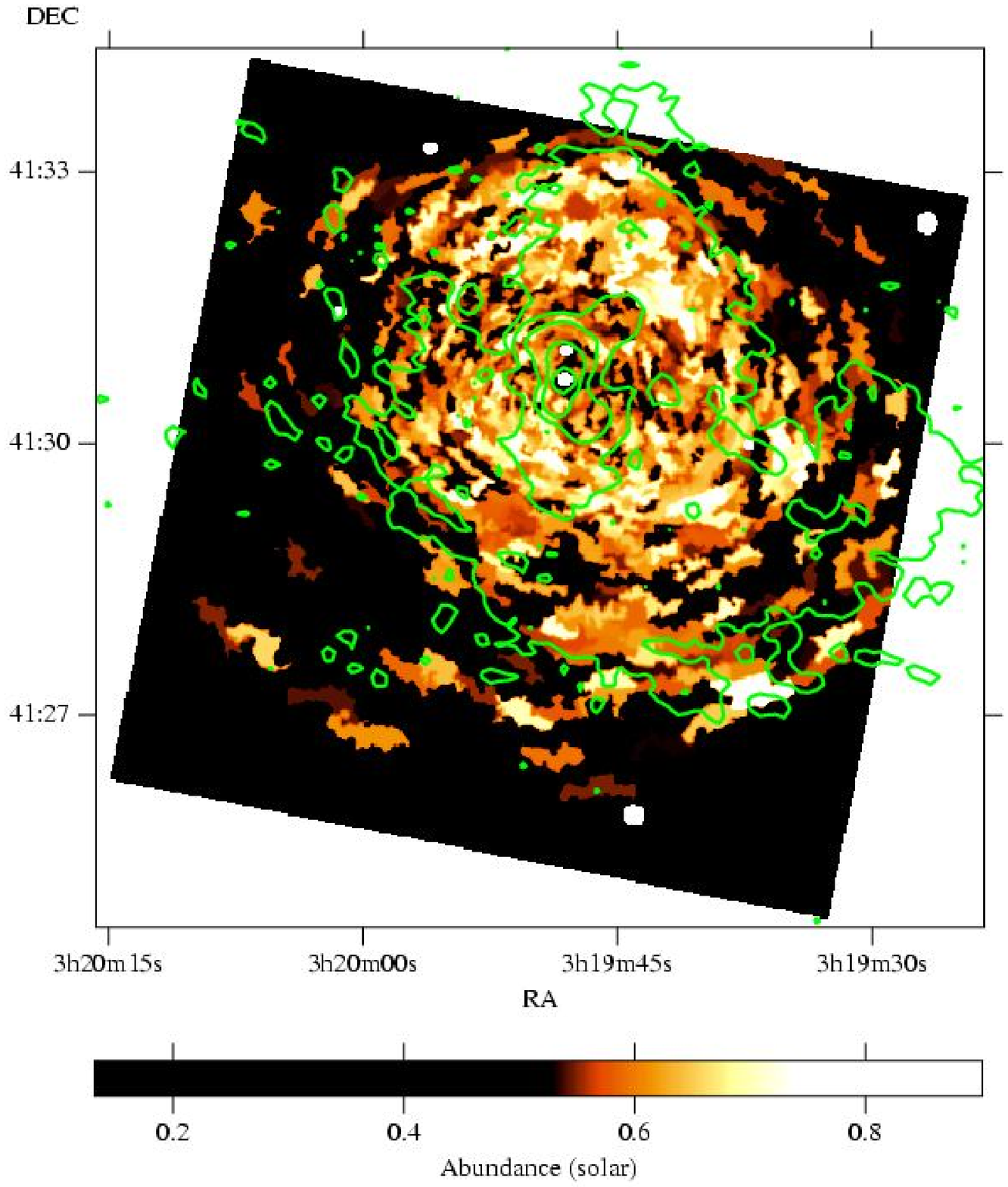}
  \includegraphics[width=\columnwidth]{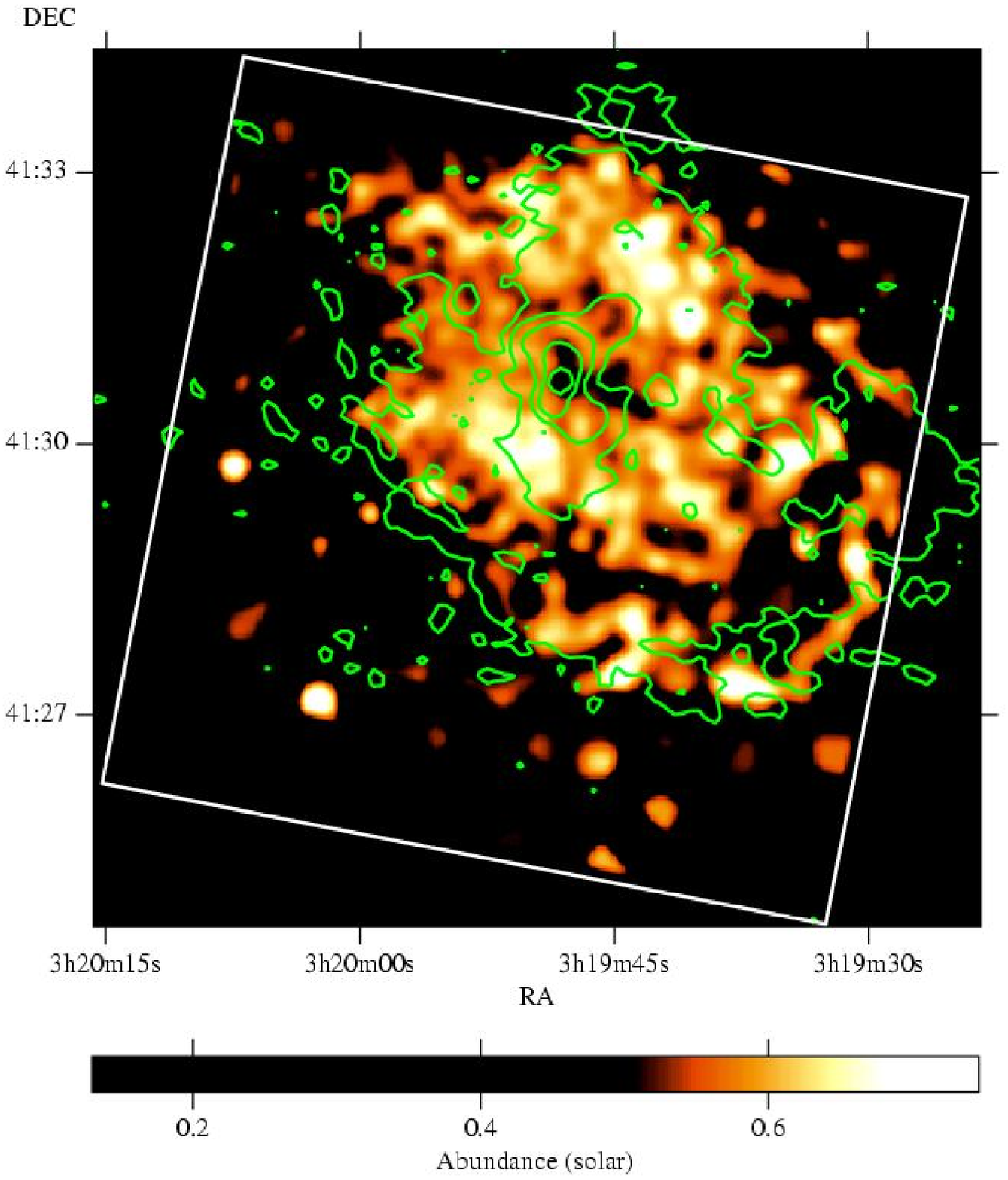}
  \caption{(Top) Abundance map of the cluster with radio contours
    overlaid.  The 1-$\sigma$ statistical uncertainties on the
    abundances range from around $0.06\Zsun$ in the centre to
    $0.1\Zsun$ at the outside.  The radio map was taken using the VLA
    in A configuration at 330~MHz for 21-ks (programme AP001). The
    radio contours are between 0.003 and 8 Jy~beam${}^{-1}$ in 6
    logarithmic steps, with a beam width of $6.25\times6.25$~arcsec.
    (Bottom) Abundance map detail using bin accretion technique. The
    uncertainty of the metallicity of each region on the edge of the
    rim is around $0.1\Zsun$. The map is smoothed with a Gaussian of
    width 6~arcsec.  The scale below each graph shows the full range
    of values in the data, but the colours have been chosen to
    highlight the high abundance shell.}
\label{fig:abundance}
\end{figure}

\begin{figure*}
  \includegraphics[width=\textwidth]{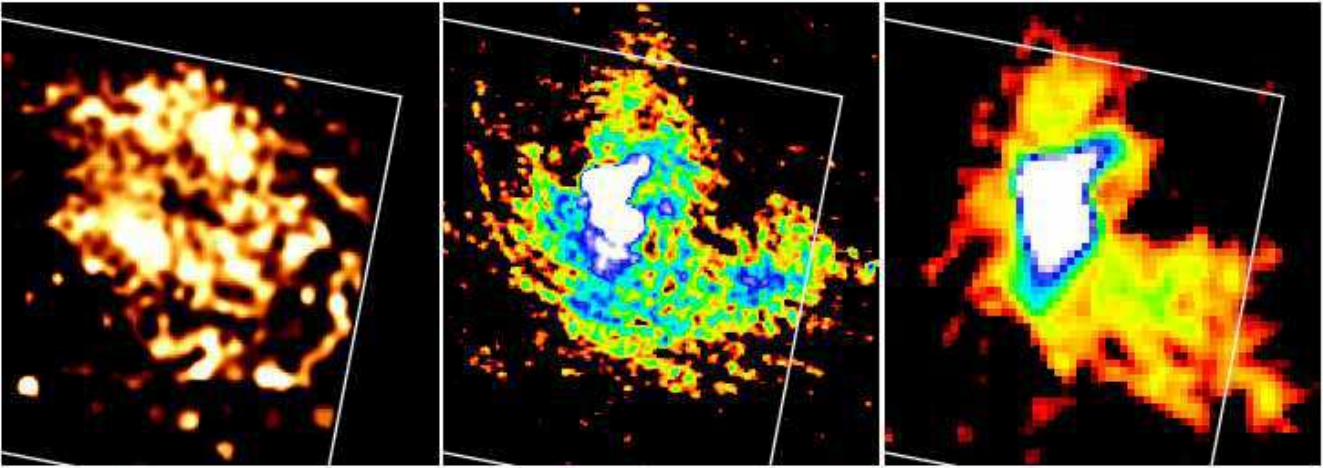}
  \caption{(Left) Abundance map created using bin accretion technique,
    smoothed with a Gaussian of width 6~arcsec. The white box shows
    the area of the CCD. The colour scale of the map is truncated to
    show only the highest metallicity regions. (Centre) 330~MHz radio
    image of the cluster. (Right) 74~MHz radio image of the same
    region (Fabian et al 2002).}
  \label{fig:radiocompar}
\end{figure*}

In Fig.~\ref{fig:temperature} we show the derived temperature map for
the cluster on the ACIS-S3 CCD. This figure can be compared to
Fig.~4~(top) in Sanders et al (2004), created using a tessellation
technique. In Fig.~\ref{fig:abundance}~(top) is shown the abundance
map. Overlaid on the abundance map are 330~MHz radio contours taken
using the VLA, showing the nuclear radio source 3C~84, its radio
lobes, the spurs pointing out towards the already known fossil radio
bubbles, and the extended mini-halo emission.

There appears to be a ridge of high abundance along the edge of the
mini-halo 250~arcsec (93~kpc) to the south-west of the nucleus.
Although the contour binning algorithm selects regions which follow
the surface brightness well, the regions created using the procedure
we described above are not as compact as those created using the bin
accretion algorithm of Cappellari \& Copin
(2003). Fig.~\ref{fig:abundance}~(bottom) shows a map made using bin
accretion, with each region containing $\sim 10^4$ counts. This map
allows us to closely identify where the metal-rich material is. We
resolve the edge of rim into a shell at the edge of the mini-halo of
maximum width $\sim 20$~arcsec (7~kpc).

In addition to the high-abundance ridge, most of the metal-rich gas in
the core of the cluster (inside a radius of $\sim 50$~kpc) is bordered
by the edge of the mini-halo. It can also be noted that the edge of
the mini-halo, where there is a change in metallicity, also
corresponds to a change in temperature (Fig.~\ref{fig:temperature}).
Another way to represent this is the inverse correlation between
temperature and abundance previously found in this cluster (Sanders et
al 2004, figures 11 and 22). In particular there is a bulge of cool
gas to the north of the core (at the edge of the map, 03:19:45,
+41:33:17), where the radio is extended
(Fig.~\ref{fig:abundance}~[bottom]; Fig.~\ref{fig:radiocompar}).

A low frequency 74~MHz image of the cluster
(Fig.~\ref{fig:radiocompar}~[right]; Fabian et al 2002) also shows a
bright region pointing towards the shell, extending beyond it.

Another noteworthy feature is the high-abundance clump at (03:19:39,
+41:29:18). This clump is seated roughly midway between the high
abundance shell and the nucleus.

\subsection{H$\alpha$ image}
\begin{figure}
  \includegraphics[width=\columnwidth]{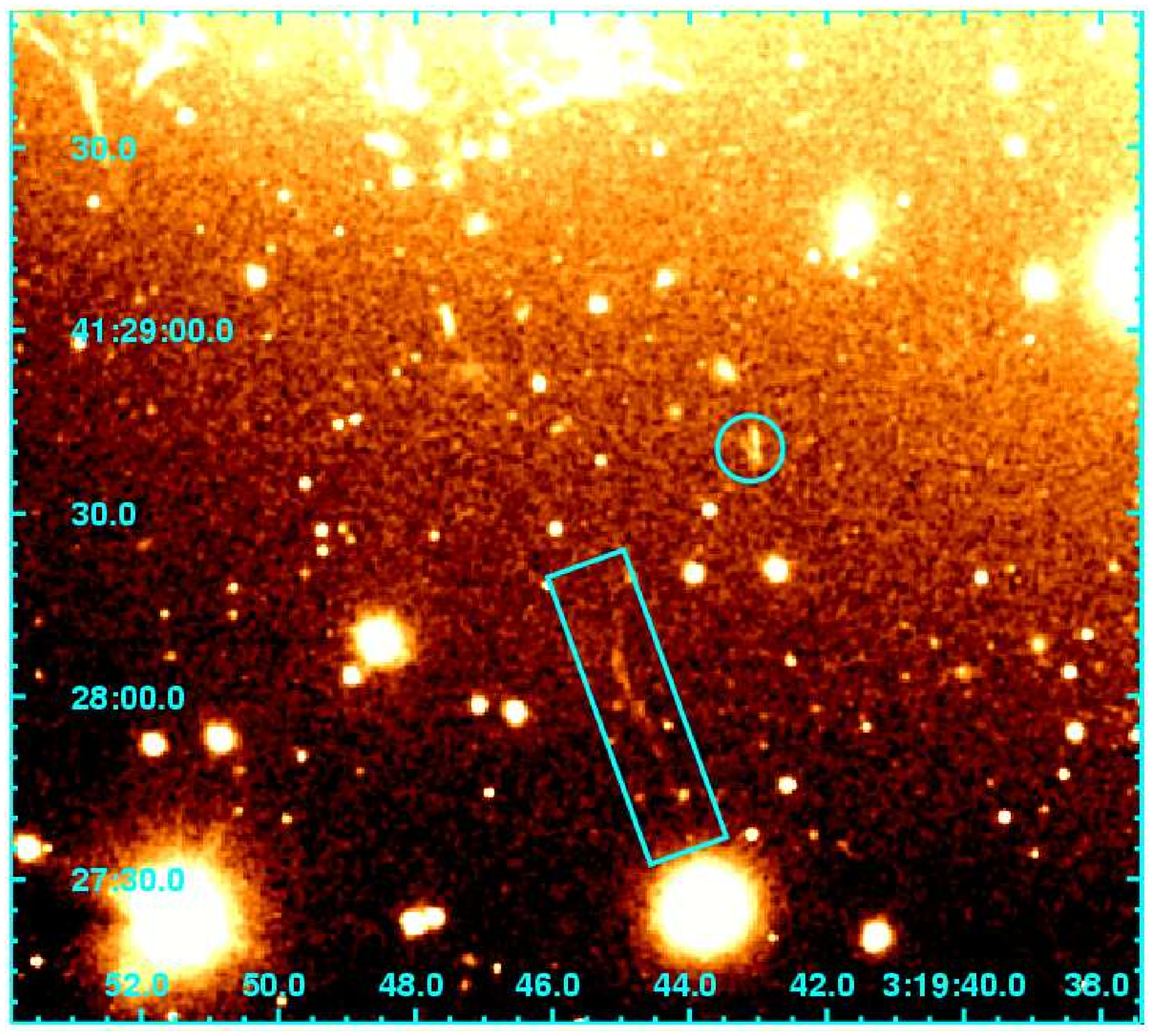}
  \caption{H$\alpha$ image of the south of the core of the Perseus
    cluster (Conselice et al 2001). This image was taken using the
    Wisconsin-Indiana-Yale-NOAO (WIYN) telescope . The large and small
    filaments are indicated with the box and circle, respectively.}
  \label{fig:halpha}
\end{figure}

We examined a H$\alpha$ map of the south of the Perseus cluster
(Conselice et al 2001), finding two previously unreported
H$\alpha$ structures to the south-south-west of the nucleus, at a
similar radius to the high abundance ridge (Fig.~\ref{fig:halpha}).
One structure is extended ($\sim 30$~arcsec, long), the other is
around 6~arcsec in length, with a V-shaped morphology. The extended
filament points from the central nucleus to the southern edge of the
high abundance shell.

\subsection{Profile across the shell}
We attempted to account for projection to investigate the physical
properties of the gas around the high abundance shell, using the
\textsc{projct} model in \textsc{xspec} to do this. Unfortunately the
results from the spectral fitting appeared to be unstable. Although we
could produce results which were plausible by choosing certain
sectors, we found that changing the position or number of the sectors
slightly resulted in temperatures which oscillated between extreme
values. The error bars on the values were incompatible with a smooth
variation. Halving the sector width doubled the frequency of the
oscillations. Therefore it is likely to be unsafe to trust chosen
sectors where the deprojection appears to work.  The likely cause for
this `temperature bouncing' is that the cluster is not spherically
symmetric over the regions examined.

\begin{figure}
  \includegraphics[width=\columnwidth]{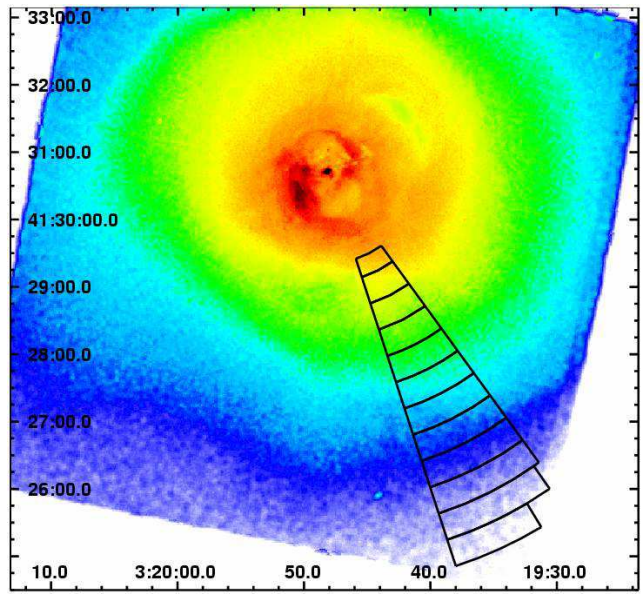}
  \caption{Smoothed full band X-ray image showing the regions used to
    generate the profiles in Fig.~\ref{fig:profile}.}
  \label{fig:profregions}
\end{figure}

\begin{figure}
  \includegraphics[width=\columnwidth]{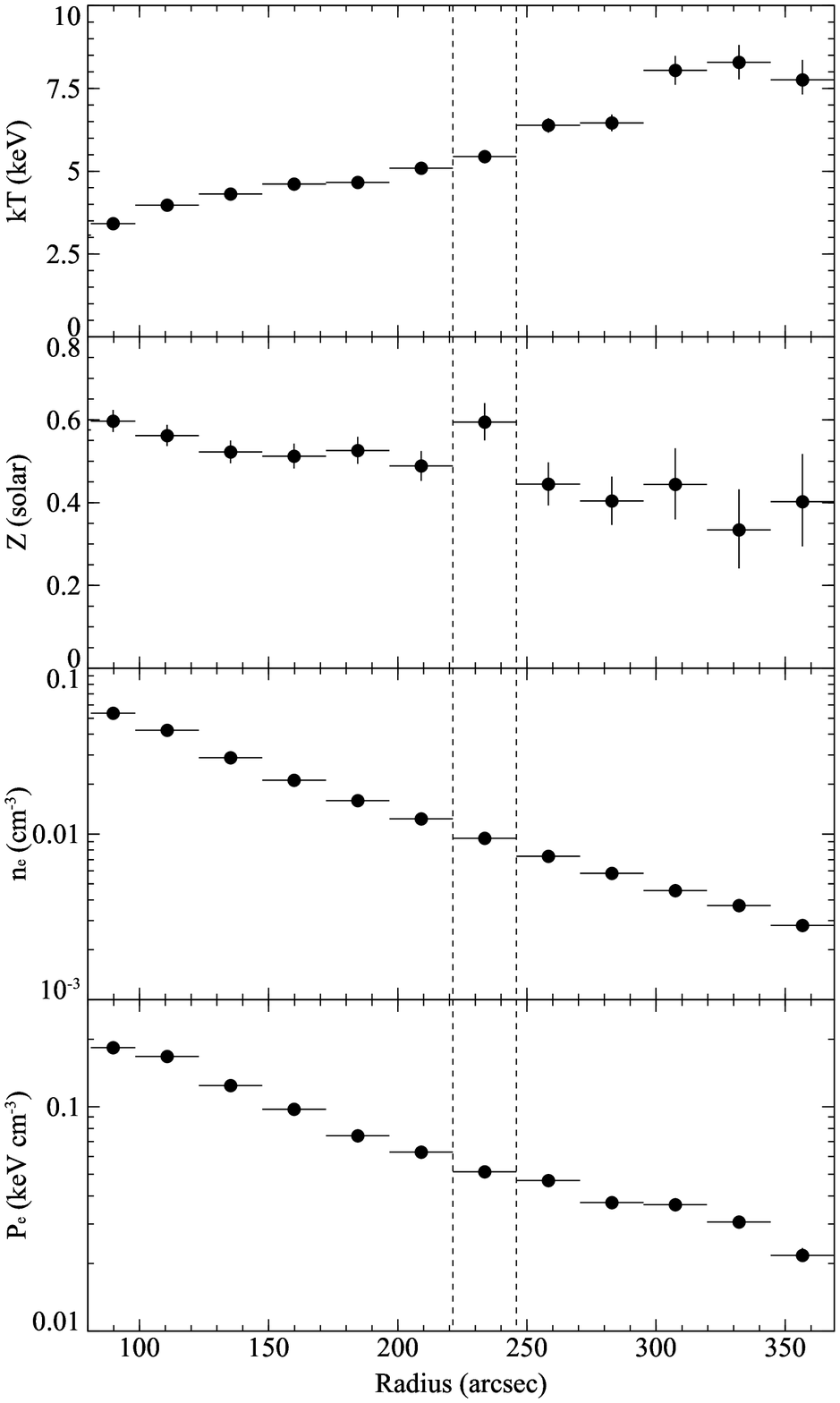}
  \caption{Projected profiles across the high abundance shell. The
    regions used are shown in Fig.~\ref{fig:profregions}. Plotted is
    the temperature, abundance, estimated electron density and
    pressure. The sector containing the high abundance shell is
    marked by dotted lines.}
  \label{fig:profile}
\end{figure}

We therefore decided not to pursue a deprojection analysis, and
instead calculated projected results for the sectors show in
Fig.~\ref{fig:profregions}. The spectrum in each sector was fitted
with a \textsc{mekal} model with the normalisation, temperature and
solar relative abundance free. Also allowed to vary in each spectral
fit was a \textsc{phabs} model to account for Galactic absorption. The
results of the spectral fits are shown in Fig.~\ref{fig:profile}. In
addition we estimated the electron density inside the shell by taking
the emission measure of the \textsc{mekal} component in the shell, and
computing the electron density assuming the volume of the shell is the
area on the sky times the radius from the centre of the cluster. The
pressure was estimated by multiplying this value by the projected
emission-weighted temperature.

The profiles show beyond the high abundance shell that there is a jump
in projected temperature by around 1~keV. However, there is no obvious
step change in electron density or pressure over this radius or
beyond. The change in temperature beyond the high abundance shell is
therefore likely to mark a cold front (Markevitch et al 2000).

\subsection{Non-thermal components}
\begin{figure}
  \includegraphics[width=\columnwidth]{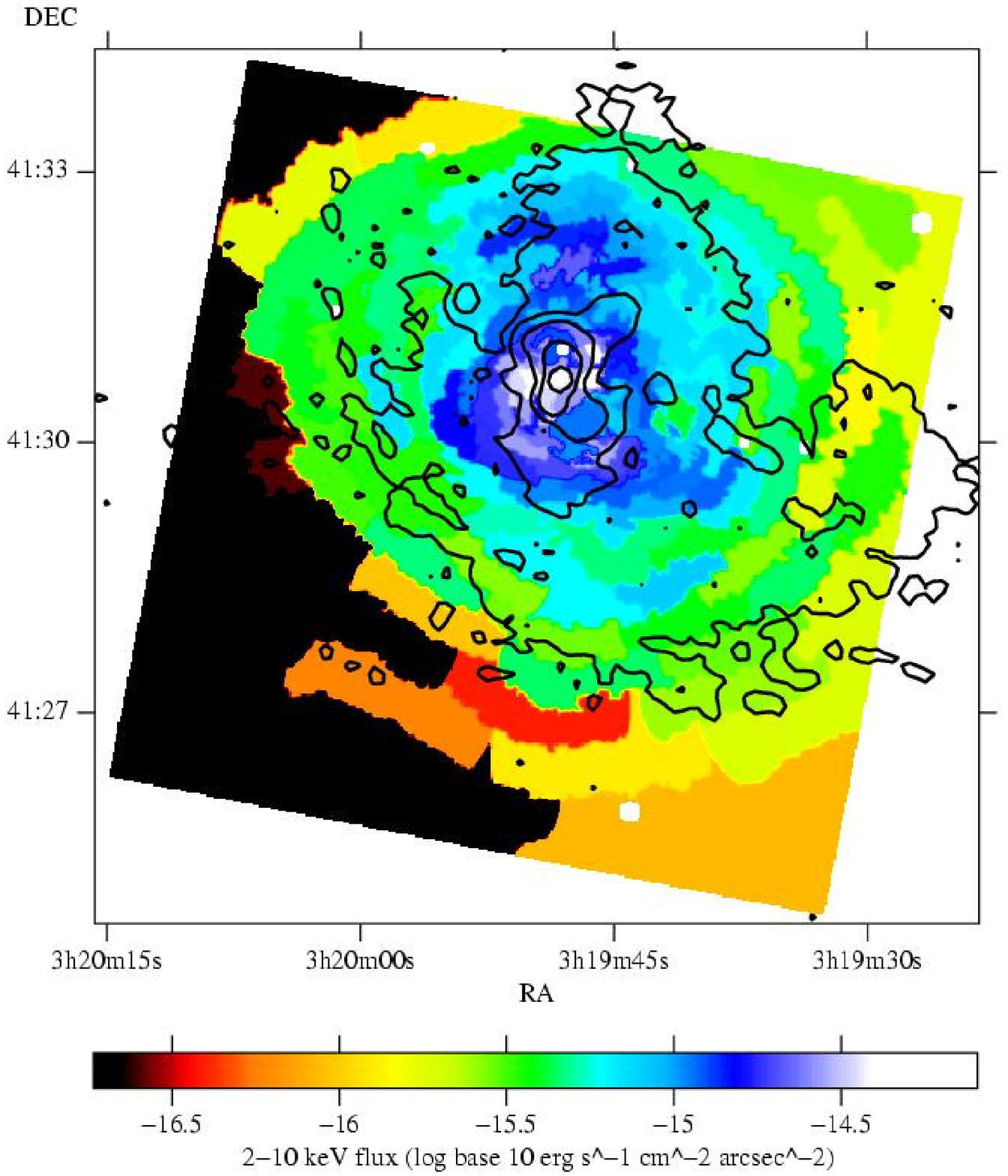}
  \includegraphics[width=\columnwidth]{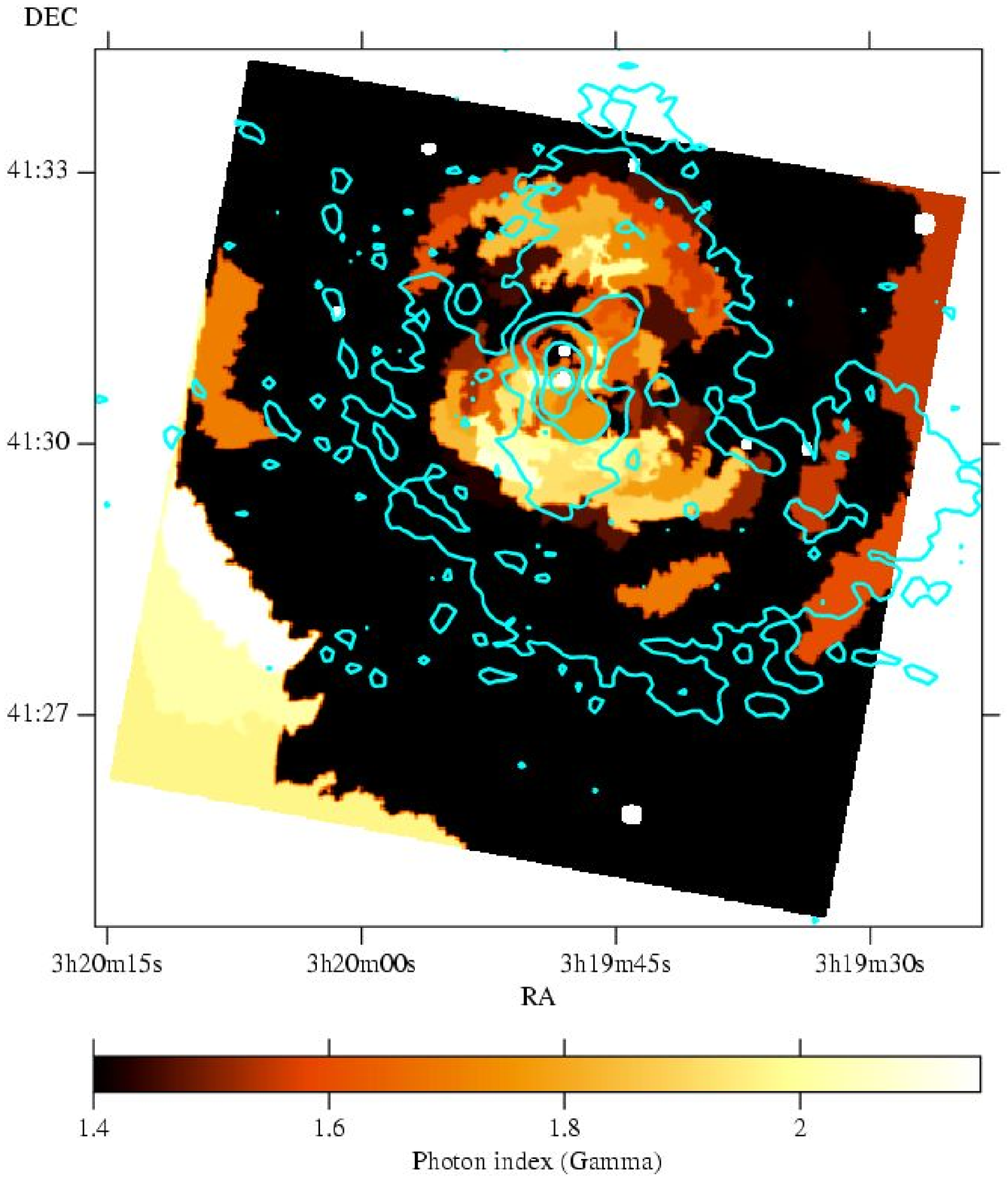}
  \caption{(Top) Powerlaw component flux (per square arcsecond in the
    2-10 keV band), and photon index (bottom). Photon indexes are
    constrained to lie between 1.4 and 2.4. The radio contours are the
    same as in Fig.~\ref{fig:abundance}.}
  \label{fig:plaw}
\end{figure}

\begin{figure}
  \includegraphics[width=\columnwidth]{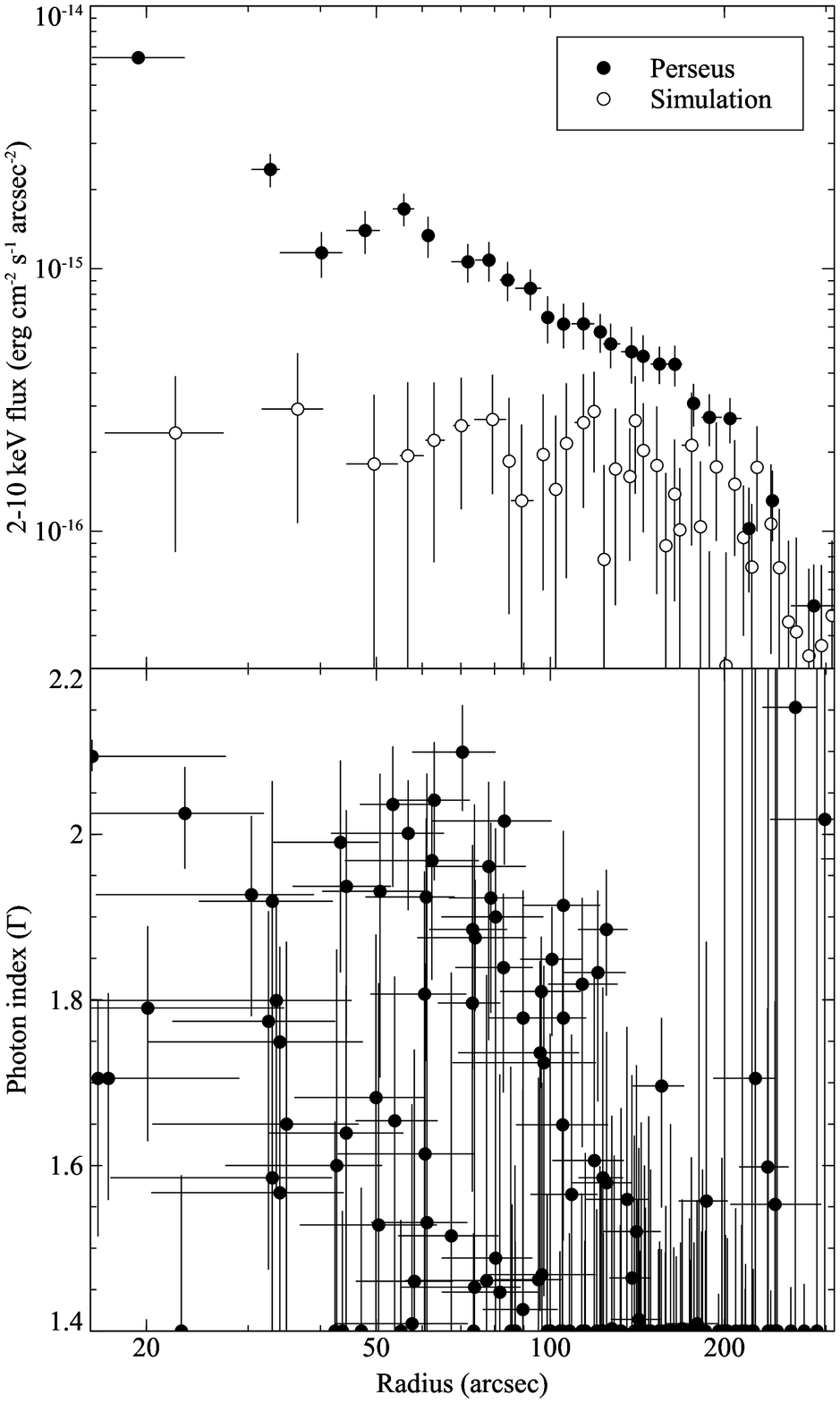}
  \caption{(Top) Weighted mean deabsorbed 2-10 keV flux of the power
    law component as a function of radius, in bins of 6 data points.
    Filled circles show the results from the real dataset, whilst
    empty circles are for the fake dataset with no intrinsic power law
    component.  (Bottom) Powerlaw index, $\Gamma$, profile of the real
    dataset.  $\Gamma$ was constrained to lie between 1.4 and 2.4.}
  \label{fig:plawprofile}
\end{figure}

We previously found evidence for the existence of a hard, probably
non-thermal, component in the X-ray spectrum from the centre of the
cluster (Fig.~15 in Sanders et al 2004), found by fitting a
high-temperature thermal component.  To understand the nature of this
component better, and to compare it against the structure of the radio
source, we have fitted a more appropriate thermal plus powerlaw model
to regions in the cluster.

We fitted spectra extracted from regions with a signal to noise ratio
of greater than 300 ($\sim 9\times 10^4$ counts) using the same
contour binning technique as described in Section~\ref{sect:temp}.
Each spectrum was fitted with a \textsc{mekal} thermal component, with
variable temperature, abundance and normalisation, plus a power-law
component, with variable photon index and normalisation, both absorbed
by a variable \textsc{phabs} absorber. The photon index of the
powerlaw component was constrained to lie between 1.4 and 2.4. We
fitted the spectra between 0.6 and 8~keV.

In Fig.~\ref{fig:plaw} is shown the X-ray flux of the powerlaw
component, in the 2-10~keV band per square arcsecond, and its photon
index, $\Gamma$. In \textsc{xspec} the powerlaw model is defined as
$A(E) = K ( E / \mathrm{keV})^{-\Gamma}$, where $K$, the
normalisation, is in units of $\mathrm{photons}\: \mathrm{keV}^{-1}\:
\mathrm{cm}^{-2}\: \mathrm{s}^{-1}$ at an energy of 1~keV.

The powerlaw flux map has a similar morphology to the hard component
map in Sanders et al (2004), but we are better able to match it to the
X-ray gas morphology with the contour binning technique. The brightest
apparent emission comes from a mushroom-shaped region to the north of
the core, to the south-east of the core, and a region pointing to the
south-west. We detect a strong component ($>
2\times10^{-16}\ergpcmsqps$~arcsec${}^{-2}$) from the inner 200~arcsec
(74~kpc) of the core, peaking at values of
$8\times10^{-15}\ergpcmsqps$~arcsec${}^{-2}$. There is also an
enhancement parallel to the high abundance ridge, just inside of it,
along where the mini-halo is extended. This enhancement is significant
statistically. The inner radio lobes are embedded in the brightest
non-thermal emission. The outer NW hole is embedded in strong emission
but shows no excess. There is a detached region of bright emission
near the southern bubble, but without exact correspondence.

The photon index of the powerlaw appears to change as a function of
position. In the regions where the component is brightest, the photon
index is large ($\sim 2.1$). This drops off in radius quickly to the
lowermost value allowed in the fit, 1.4.

We note that this two-dimensional mapping of the powerlaw component
does not include the effects of projection.  It is expected there will
be overlying hot gas of around 7~keV in front of the cool region. This
will contribute to the non-thermal component detected. In order to
check whether projection affects the results, we simulated a 200~ks
\emph{Chandra} observation of the Perseus cluster with no powerlaw
component, to repeat the analysis. We took the radial temperature and
density profile parameterisations given in Churazov et al (2003).
These have the advantage that they go out to larger radii than our
\emph{Chandra} measurements. We also took a simple cubic fit to the
deprojected abundance profile of Sanders et al (2004), truncating it
at $0.3\Zsun$ above 120~kpc. Using these profiles, we simulated a
$\sim 600 \times 600 \times 1000 \: \mathrm{arcsec}^3$ volume ($\sim
220 \times 220 \times 360 \kpc^3$; where the $z$ direction is along
the line of sight) of the cluster.  In regions of $\sim 4 \times 4
\times 8$~arcsec we generated a simulated spectrum for the plasma at
that radius using \textsc{mekal}, \textsc{phabs} and \textsc{xspec}.
We extracted the photons which made up each spectrum, randomising
their position on the cuboid projected on the sky. Using these
photons, we populated an event file suitable for analysis with the
\textsc{ciao} tools. We added X-ray background photons from a faked
spectrum generated using a three-powerlaw fit to the \emph{Chandra}
blank-sky background spectrum.

In Fig.~\ref{fig:plawprofile}~(top), a radial profile shows the
average measured powerlaw flux per square arcsecond. Also plotted is
the measured value from the analysis of the simulated dataset, which
does not include any non-thermal emission.  Although there is a weak
signal from projection effects, the observed signal is over an order
of magnitude larger than the background in the centre. There appears
to be an excess out to radii of at least 100~arcsec, and probably
200~arcsec.

Another potential source for the excess emission are the Ni and Fe-K
lines. We tried to use a model with variable Ni abundance, but it did
not appear to significantly change the powerlaw normalisation in the
innermost region.  X-ray background effects remain a slight
possibility, although it is unclear why there would be an order of
magnitude difference in the centre of the field from the outer
regions.

The photon index is well constrained at around 2 for the brightest
region of emission (Fig.~\ref{fig:plawprofile}, bottom). The photon
index measurements are largely unconstrained from the simulated dataset.

The total powerlaw flux is $6.3 \times 10^{-11} \ergpcmsqps$ between 2
and 10~keV, found by integrating Fig.~\ref{fig:plawprofile}~(top), and
subtracting the flux found using the simulated dataset. This
corresponds to a luminosity of $4.8 \times 10^{43} \ergps$, which is
similar to the luminosity given by Sanders et al~(2004).

\subsubsection{Estimating the magnetic field}
If the non-thermal emission is the result of inverse Compton emission,
it is possible to estimate the value of the magnetic field. If there
is a photon field with energy density $E_\mathrm{ph}$, the ratio of
the non-thermal X-ray flux to radio flux is approximately
\begin{equation}
  \frac{L_X}{L_R} = \frac{E_\mathrm{ph}}{B^2/8\pi} 
  \left( \frac{\gamma_X}{\gamma_R} \right)^2
  \frac{N(\gamma_X)}{N(\gamma_{R})}
  = \frac{E_\mathrm{ph}}{B^2/8\pi} \left(  \frac{\gamma_X}{\gamma_R} \right)^
  {1-2\alpha},
\end{equation}
where $N(\gamma)$ is the number density of electrons with Lorentz
factor $\gamma$, $\gamma_X = \left( \nu_X / \nu_\mathrm{orig}
\right)^{1/2}$, $\nu_X$ and $\nu_\mathrm{orig}$ are the frequencies of
the X-ray and the radiation which is scattered, $\gamma_R = \left(
  \nu_R/\nu_\mathrm{cyc} \right)^{1/2}$, the cyclotron frequency
$\nu_\mathrm{cyc} / \mathrm{Hz} \sim 4 \times 10^6 (B / \mathrm{G})$,
$B$ is the magnetic field and $\nu_R$ is the frequency of the measured
radio flux.

There are two dominant sources of photons for inverse Compton
scattering in this object. These are the Cosmic Microwave Background
(CMB), and the infrared (IR) flux from NGC~1275 ($1.6\times10^{11}
\Lsun$; Impey \& Neugebauer 1988). The non-thermal flux observed will
be the sum of these two contributions,
\begin{eqnarray}
  \frac{L_X}{L_R} = \frac{1 }{B^2 / 8\pi} 
  \left( \frac{1}{\gamma_R} \right)^{1-2\alpha}
  \left[
    E_\mathrm{CMB}
    \left(\frac{\nu_X}{\nu_\mathrm{CMB}}\right)^{(1-2\alpha)/2}+
    \right. \nonumber \\
    \left. E_\mathrm{IR}
    \left(\frac{\nu_X}{\nu_\mathrm{IR}}\right)^{(1-2\alpha)/2}
  \right],
\end{eqnarray}
where $\nu_\mathrm{CMB}$ is the frequency of the CMB radiation ($\sim
160$~GHz), $\nu_\mathrm{IR}$ is the frequency of the IR radiation
(equivalent to $100 \mu$m), $E_\mathrm{CMB}$ is the energy density of
the CMB field ($4.5 \times 10^{-13}\ergpcmcu$), and $E_\mathrm{IR} =
L_\mathrm{IR} / (4\pi c R^2)$ at a radius of $R$.  This equation can
be solved for $B$ by substituting in $\gamma_R$.

\begin{figure}
  \includegraphics[width=\columnwidth]{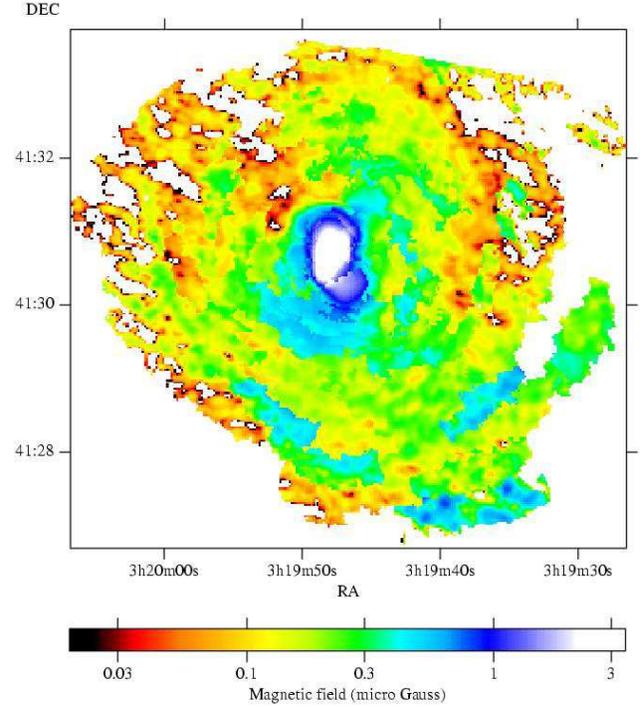}
  \caption{Estimated magnetic field over the core of the cluster.
    There is no signal in the white outer parts of the images. Values
    of $B$ below $0.1\mu$G are unlikely to be accurate.}
  \label{fig:bmap}
\end{figure}

Taking the 330~MHz radio map of the mini halo and dividing it by the
area of the beam, we computed the radio flux per square arcsecond.
Using this map with the X-ray powerlaw flux (per square arcsecond) and
photon index maps, we estimated the magnetic field as a function of
position.  A value of $2.4 \times 10^{-16}\ergpcmsqps$~arcsec${}^{-2}$
was subtracted from the X-ray map to account for the ``projection
background'' shown in Fig.~\ref{fig:plawprofile}.  In
Fig.~\ref{fig:bmap} is shown the estimated magnetic field over the
core of the cluster.  The contribution from the IR photon field
dominates the CMB contribution within a radius of around $40\kpc$.

At the edge of the field we are close to the limit of the radio image.
Therefore the values of $B$ are unlikely to be accurate. Increased
values of $\Gamma$ would also raise our estimates of the magnetic
field there too.  Furthermore the uncertainties of the subtraction of
the background signal from the X-ray flux are much more significant
near the edge of the field.  We therefore recommend that values of $B$
below $0.1\mu$G are ignored.

The calculated values will also be affected if the magnetic field is
filamentary. The synchrotron emission could originate from regions
with a different magnetic field from the inverse Compton emission.
Highly filamentary polarisation structure has been seen from the radio
galaxy Fornax A (Fomalont et al 1989).

We note that the inferred pressure of the relativistic electrons which
would produce the observed non-thermal emission approaches 60~per~cent
of the total pressure in the small regions where $\Gamma$ is high
($\gtrsim 2$) to the south and north of the nucleus (see
Fig.~\ref{fig:plaw}~[bottom]). Elsewhere it ranges from 1 to
30~per~cent of the total pressure, declining quickly with radius. If
this result is confirmed by deeper observations of Perseus, it would
have consequences for heating in clusters.

\section{Discussion}
\subsection{High abundance ridge}
The ridge appears not to be dependent on the binning technique used,
which suggests the feature is robust. If the ridge is a real feature,
it may be high abundance material lifted out of the inner core,
material deposited in a merger event, or it may have formed in situ.

It seems unlikely that the metals were deposited by stars in the
current location of the ridge. The high abundance region is a large
distance from the core of NGC~1275. A merger remains a possibility. If
this is the case it may have also disturbed the radio morphology to
match the location of the high abundance material. It is however
difficult to interpret the radial H$\alpha$ filaments in this
scenario.

Our favoured explanation is that the material was lifted from the core
of the cluster. There are correspondences between the outer north-west
and inner south-west radio lobe positions and metallicity
enhancements. Simulations of buoyantly rising bubbles in clusters show
that material is entrained by the rising bubble (Churazov et al 2001,
see figure 9). Gas is entrained at the upper surface of the rising
bubble, and in its wake. The ridge of high abundance material could
represent the upper (or lower) surface of a rising bubble. The high
abundance clump midway between the core of the cluster and the ridge
could be material lifted in the wake of the bubble.

This explanation also fits in naturally with the extended H$\alpha$
filament which is pointing towards the south of the ridge. Previously
two filaments behind the north-west bubble were identified, which
appear to be acting as streamlines (Fabian et al 2003b). If this is
the case the filaments trace the motion of the intracluster medium,
the flow is laminar, and the gas here is not turbulent. Models of
rising bubbles indicate that viscosity can suppress the instabilities
that lead to the shredding of rising bubbles (Reynolds et al 2004).
Furthermore sound waves generated by the expanding bubble may
viscously heat the core of the cluster, thereby offsetting cooling
(Fabian et al 2003a).

Further evidence is suggested by the correlation between the high
abundance ridge and the edge of the mini-halo. A bubble will detach
and rise until the density of its contents matches the density of the
surrounding gas. If viscosity and magnetic field effects are
negligible, instabilities will destroy the bubble before it reaches
this radius. If this is not the case, the bubble will then flatten and
grow around the isodensity surface (forming a ``pancake''), and remain
intact as long as surface effects retain its structure.  For each of
the existing bubbles in Perseus, we find associated radio emission.
The mini-halo may be the radio counterpart to the bubble that formed
the ridge.  The ridge lies at a temperature and density interface, and
so is a natural interface where a bubble would pancake. The
non-thermal X-ray emission associated with the rim suggests the
presence of old electrons which made up the bubble.

It may be the case that the halo is the remaining low frequency
emission of all the old radio bubbles in the
cluster\footnote{Alternatively, Gitti et al (2002) have proposed that
the radio mini-halo is due to turbulent reacceleration of intracluster
cosmic-ray electrons.}. This would be the case if the bubbles were
always generated with similar densities.  This idea may fit with the
inhomogeneous metallicity map of the core of the cluster
(Fig.~\ref{fig:abundance}).  Rising bubbles may be responsible for
much of the structure in this image, displacing metal-rich gas from
the core to larger radii.  Then there is very little mixing taking
place in the intracluster medium, with the gas viscous and not
turbulent.  The width of the high abundance ridge ($\sim 20$~arcsec;
7~kpc) places limits on mixing. We can estimate ages for the bubble,
following the approach of Dunn \& Fabian (2004), of $10^8$ (rising at
buoyancy velocity), and $9 \times 10^7$~yr (refilling velocity). These
are the likely timescales for which the ridge must have survived if it
were at the edge of a bubble. See Dunn \& Fabian (2004) and Dunn \&
Fabian (in preparation) for an explanation of these timescales and the
assumptions made in their calculation.

There are some possible difficulties with this model. We do not know
whether it is possible to entrain enough metals with a rising bubble,
especially on its rim. We do not know how entrainment is affected by
viscosity and magnetic fields. Nevertheless, we observe high
abundances patches around at least two of the existing radio bubbles.
We do not know whether the ICM is viscous or turbulent.  Future
observations of Perseus using \emph{ASTRO-E2} should resolve this
issue. Indeed, forthcoming analysis of deeper observations of Perseus
by \emph{Chandra} will enable us to map the abundance in exquisite
detail. In addition we will be able to confirm and further examine the
bulk motion found in the core of the cluster (Sanders et al 2004).

We presume that the high metallicity shell with break up and the
iron-rich denser material will fall back towards the centre.
Otherwise, it is difficult to understand how the central regions have
remained at high abundance. This will further heat the inner regions.
Flows may therefore take place in both directions.

\subsection{Non-thermal emission}
The evidence for non-thermal emission may still be an artifact of the
spectral fitting procedure.  Unfortunately, since the
\emph{XMM-Newton} observation of this cluster is affected by high
background (Churazov et al 2003), we were unable to easily confirm the
non-thermal X-ray emission with this instrument. It is therefore
important to observe this cluster with \emph{XMM-Newton} again in a
period with low background.

If our identification of non-thermal emission is correct, then it
opens the possibility of real detections of inverse Compton emission
in other clusters of galaxies by current X-ray telescopes. Measuring
magnetic fields by this method would complement existing methods (see
Carilli \& Taylor 2002).

\subsection{Further fossil bubbles}
There is a further interesting connection between the abundance,
temperatures, radio, H$\alpha$ and non-thermal maps.  All of these
maps show features pointing towards the north. These include the bulge
at the top of the temperature map (Fig.~\ref{fig:temperature}), the
extension of the 330 and 74~MHz radio images in that direction
(Fig.~\ref{fig:radiocompar}), the long H$\alpha$ filaments pointing
north (Conselice et al 2001) and the extension of the non-thermal
component in that direction (Fig.~\ref{fig:plaw}). This may indicate
that this is the path of a previous buoyant bubble.  Furthermore there
are further structures in the metallicity and non-thermal maps which
may be the result of other fossil bubbles.

\section*{Acknowledgements}
ACF and RJHD thank the Royal Society and PPARC for support,
respectively.

\end{document}